\input harvmac
\noblackbox
\newcount\figno
\figno=0
\def\fig#1#2#3{
\par\begingroup\parindent=0pt\leftskip=1cm\rightskip=1cm\parindent=0pt
\baselineskip=11pt
\global\advance\figno by 1
\midinsert
\epsfxsize=#3
\centerline{\epsfbox{#2}}
\vskip 12pt
\centerline{{\bf Figure \the\figno:} #1}\par
\endinsert\endgroup\par}
\def\figlabel#1{\xdef#1{\the\figno}}
\def\pano{\par\noindent}

\def\meno{\medskip\noindent}

\font\cmss=cmss10
\font\cmsss=cmss10 at 7pt
\def\rlx{\relax\leavevmode}
\def\inbar{\vrule height1.5ex width.4pt depth0pt}
\def\IC{\relax\,\hbox{$\inbar\kern-.3em{\rm C}$}}
\def\IN{\relax{\rm I\kern-.18em N}}
\def\IP{\relax{\rm I\kern-.18em P}}
\def\ZZ{\rlx\leavevmode\ifmmode\mathchoice{\hbox{\cmss Z\kern-.4em Z}}
 {\hbox{\cmss Z\kern-.4em Z}}{\lower.9pt\hbox{\cmsss Z\kern-.36em Z}}
 {\lower1.2pt\hbox{\cmsss Z\kern-.36em Z}}\else{\cmss Z\kern-.4em Z}\fi}
\def\narrowplus{\kern -.04truein + \kern -.03truein}
\def\narrowminus{- \kern -.04truein}
\def\narrowminussub{\kern -.02truein - \kern -.01truein}

\def\cl{\centerline}

\def\sf{{\cal F}}
\def\sg{{\cal G}}
\def\sh{{\cal H}}
\def\sm{{\cal M}}
\def\sV{{\cal V}}
\def\HH{{\rm I\hskip -2pt H\hskip -2pt I}}

\def\sqr#1#2{{\vcenter{\vbox{\hrule height.#2pt
            \hbox{\vrule width.#2pt height#1pt \kern#1pt
                  \vrule width.#2pt}\hrule height.#2pt}}}}


\lref\rbanks{ T.\ Banks, L.J.\ Dixon, D.\ Friedan and E.\ Martinec,
  {\it Phenomenology and conformal field theory, or Can string theory predict
  the weak mixing angle}?,
   Nucl.\ Phys.\ {\bf B299} (1988) 613}

\lref\rbsw{R.\ Blumenhagen and A.\ Wi{\ss}kirchen,
  {\it Exactly solvable $(0,2)$ supersymmetric string vacua with GUT
  gauge groups},
  Nucl.\ Phys.\ {\bf B454} (1995) 561, hep--th/9506104\semi
 R.\ Blumenhagen, R.\ Schimmrigk and A.\ Wi{\ss}kirchen,
  {\it The $(0,2)$ exactly solvable structure of chiral rings,
  Landau--Ginzburg theories and Calabi--Yau manifolds},
  Nucl.\ Phys.\ {\bf B461} (1996) 460, hep--th/9510055 \semi
  R.\ Blumenhagen and A.\ Wi{\ss}kirchen,
  {\it Exploring the moduli space of $\,(0,2)$ strings},
  Nucl.\ Phys.\ {\bf B475} (1996) 225, hep-th/9604140}

\lref\rbsf{R.\ Blumenhagen, R.\ Schimmrigk and A.\ Wi{\ss}kirchen,
  {\it $(0,2)$ mirror symmetry},
  Nucl.\ Phys.\ {\bf B486} (1997) 598, hep--th/9609167 \semi
  R.\ Blumenhagen and S.\ Sethi,
  {\it On orbifolds of $(0,2)$ models},
  Nucl.\ Phys.\ {\bf B491} (1997) 263, hep--th/9611172 \semi 
  R.\ Blumenhagen and M.\ Flohr,
  {\it Aspects of  $(0,2)$ orbifolds and mirror symmetry},
  IASSNS-HEP-97/13,  hep--th/9702199}

\lref\rtdual{ R.\ Blumenhagen,
     {\it Target space duality for $(0,2)$ compactifications},
   IASSNS-HEP-97/86,  hep--th/9707198}

\lref\rduala{ J.\  Distler and S.\ Kachru,
     {\it Duality of $(0,2)$ String Vacua},
     Nucl.\ Phys.\ {\bf B442} (1995) 64, hep-th/9501111}

\lref\rdualb{ T.-M.\ Chiang, J.\ Distler, B.\ R.\  Greene,
  {\it Some Features of $(0,2)$ Moduli Space},
       Nucl.\ Phys.\  {\bf B496} (1997) 590, hep-th/9702030}

\lref\rend{J.\  Distler, B.\ R.\ Greene, K.\ Kirklin and
           P.\ Miron,
    {\it Calculating endomorphism valued cohomology:
         Singlet spectrum in superstring models},
    Commun.\ Math.\ Phys.\ {\bf 122} (1989) 117\semi
    M.\ G.\ Eastwood and T.\ H\"ubsch,
    {\it Endomorphism valued cohomology and
         gauge neutral matter},
    Commun.\ Math.\ Phys.\ {\bf 132} (1990) 383\semi
    T.\ H\"ubsch, 
    {\it Calabi-Yau manifolds},
    World Scientific 1992}

\lref\rgrif{P.\ Griffiths and J.Harris,
   {\it Principles of algebraic geometry},
   John Willey \& Sons 1978\semi
     R.\ Hartshorne,
    {\it Algebraic geometry},
    Springer-Verlag, 1977\semi
      C.\ Okonek, M.\ Schneider and H.\ Spindler,
    {\it Vector bundles on complex projective spaces},
     Birkh\"auser 1980 }

\lref\rktb{ P.\ Berglund, C.\ V.\ Johnson, S.\ Kachru and 
            P.\ Zaugg,
       {\it Heterotic coset models and $(0,2)$ string vacua},
        Nucl.\ Phys.\ {\bf B460} (1996) 252, hep-th/9509170\semi
       M.\ Kreuzer and M.\ Nikbakht-Tehrani,
      {\it $(0,2)$ string compactifications},
      Talk at the International Symposium on the Theory
       of Elementary Particles, Buckow, August 27-31,
      hep-th/9611130 }

\lref\rsharpe{E.\ Sharpe,
        {\it Extremal transitions in heterotic string theory},
       PUPT-1707, hep--th/9705210} 

\lref\rew{E.\ Witten,
  {\it Phases of $N=2$ theories in two dimensions},
  Nucl.\ Phys.\ {\bf B403} (1993) 159, hep--th/9301042}

\lref\rkw{S.\ Kachru and E.\ Witten,
  {\it Computing the complete massless spectrum of a Landau--Ginzburg
  orbifold},
  Nucl.\ Phys.\ {\bf B407} (1993) 637, hep--th/9307038}

\lref\rdk{J.\ Distler and S.\ Kachru,
  {\it $(0,2)$ Landau--Ginzburg theory},
  Nucl.\ Phys.\ {\bf B413} (1994) 213, hep--th/9309110\semi
  J.\ Distler,
  {\it Notes on $(0,2)$ superconformal field theories},
   in Trieste HEP Cosmology 1994,hep--th/9502012}

\lref\rvitja{P.\ S.\ Aspinwall, B.\ R.\ Greene and D.\ R.\ Morrison,
    {\it Calabi-Yau moduli space, mirror manifolds and spacetime
     topology change in string theory},
    Nucl.\ Phys.\ {\bf B416} (1994) 414, hep--th/9309097\semi
  V.\ V.\ Batyrev,
  {\it Dual polyhedra and mirror symmetry for Calabi--Yau hypersurfaces
  in toric varieties},
  J.\ Alg.\ Geom.\ {\bf 3} (1994) 493, alg--geom/9310003}

\lref\rztwoinst{E.\ Witten,
{\it New issues in manifolds of $SU(3)$ holonomy},
   Nucl.\ Phys.\ {\bf B268} (1986) 79\semi
  M.\ Dine, N.\ Seiberg, X.\ Wen and E.\ Witten,
  {\it Nonperturbative effects on the string world sheet I,II},
  Nucl.\ Phys.\ {\bf B278} (1986) 769;
  {\it ibid.}\ {\bf B289} (1987) 319\semi
 J.\ Distler,
  {\it Resurrecting $(2,0)$ compactifications},
  Phys.\ Lett.\ {\bf B188} (1987) 431}

\lref\rdistgr{ J.\ Distler and B.\ R.\ Greene,
  {\it Aspects of $\,(2,0)$ string compactifications},
  Nucl.\ Phys.\ {\bf B304} (1988) 1}

\lref\rresolv{J.\ Distler, B.\ R.\ Greene and D.\ R.\ Morrison,
  {\it Resolving singularities in $(0,2)$ models},
   Nucl.\ Phys.\ {\bf B481} (1996) 289, hep--th/9605222\semi
   J.\ Distler, {\it private communication}}

\lref\rpri{J.\ Distler and S.\ Kachru,
  {\it Singlet couplings and $(0,2)$ models},
  Nucl.\ Phys.\ {\bf B430} (1994) 13, hep--th/9406090\semi
 E.\ Silverstein and E.\ Witten,
  {\it Criteria for conformal invariance of $\,(0,2)$ models},
   Nucl.\ Phys.\ {\bf B444} (1995) 161, hep--th/9503212}

\lref\rft{S.\ Sethi, C.\ Vafa, E.\ Witten,
      {\it Constraints on low-dimensional string compactifications},
      Nucl.\ Phys.\ {\bf B480} (1996) 213-224, hep-th/9606122 \semi
      I.\ Brunner, M.\ Lynker and R.\ Schimmrigk,\
      {\it F-theory on Calabi-Yau fourfolds},
     Phys.\ Lett.\ {\bf B387} (1996) 750, hep-th/9606148 \semi
    P.\ Mayr,
    {\it Mirror symmetry, $N=1$ superpotentials and tensionless 
     strings on Calabi-Yau  four-folds},
     Nucl.\ Phys.\ {\bf B494} (1997) 489, hep-th/9610162 \semi
    R.\ Friedman, J.\ Morgan and E.\ Witten,
    {\it Vector bundle and F-theory},
    IASSNS-HEP-97-3,  hep-th/9701162 \semi
    M.\ Bershadsky, A.\ Johansen, T.\ Pantev and V.\ Sadov,
    {\it Four dimensional compactification of F-theory},
    HUTP-96-A054, hep-th/9701165}

\lref\rschu{ S.\ Katz and S.\ A.\ Str\o mme,
         {\it Schubert: a maple package for intersection
       theory}\semi
        J.\ DeLoera,
      {\it Puntos} }
     
\Title{\vbox{\hbox{hep--th/9710021}
                 \hbox{IASSNS--HEP--97/107}}}
{\vbox{ \hbox{$(0,2)$ Target Space Duality, CICYs} \vskip 0.5cm
        \hbox{\phantom{nnnnn}and Reflexive Sheaves }}}
\smallskip
\centerline{{Ralph Blumenhagen${}^1$}    }
\bigskip
\centerline{${}^{1}$ \it School of Natural Sciences,
                       Institute for Advanced Study,}
\centerline{\it Olden Lane, Princeton NJ 08540, USA}
\smallskip
\bigskip
\bigskip\bigskip
\centerline{\bf Abstract}
\noindent
It is shown that the recently proposed target space duality
for $(0,2)$ models is not limited to models admitting a 
Landau-Ginzburg description. By studying some generic  examples
it is established for the broader class of vector bundles 
over complete intersections in toric varieties. 
Instead of sharing a common Landau-Ginzburg locus, a pair of dual
models agrees in more general non-geometric phases.  
The mathematical tools for treating
reflexive sheaves are  provided, as well.

\footnote{}
{\pano
${}^1$ e--mail:\ blumenha@sns.ias.edu
\pano}
\Date{10/97}
\newsec{Introduction}

The existence of chiral matter in nature leads one to consider
string unifications with $N=1$ supersymmetry in four space time
dimensions.  
For the heterotic string  the condition of $N=1$ supersymmetry implies  
$(0,2)$ supersymmetry on the string world sheet \rbanks.
This class of models has been the subject of study during the last 
decade \refs{\rztwoinst\rdistgr\rew\rdk\rduala\rpri\rbsw\rresolv\rdualb
\rktb{--}\rbsf}.
Besides the non-perturbative duality to F-theory vacua \rft, one recent 
development in the study of $(0,2)$ models was the appearance 
of a perturbative duality first introduced in \rduala, saying that
at large distance completely different looking $(0,2)$ compactifications
can share the same Landau-Ginzburg locus.   
In \rtdual\  it was argued that analogous to mirror symmetry this duality
is promoted to a target space duality in the sense that the entire 
perturbative moduli
spaces of a dual pair are isomorphic. The main ingredient for this conclusion
was an exact computation of the total dimension of both geometric
moduli spaces, in 
particular including bundle deformations. It turned out, that untouched 
by a possibly enlarged moduli space at the Landau-Ginzburg locus, the numbers
of large radius moduli agree. 

The future practical usefulness of this duality crucially depends on how far 
one will be able to nail down the exact relation between the moduli of a 
dual pair. 
This is a difficult question, logically similar to finding the exact
form of the mirror map for Calabi-Yau threefolds. 
In this paper we want to  address a  more moderate  question.
Namely to which extend 
$(0,2)$ target space duality depends on the special feature of an existing 
Landau-Ginzburg phase or whether it is a much more general structure
for, in general, reflexive sheaves over complete intersections in
toric varieties. 

As will be verified in section 3, the restriction to Landau-Ginzburg models
was  rather historically motivated and target space duality extends to 
much more general $(0,2)$ models as defined by vector bundles 
over complete intersections in toric varieties in the first place.  
Furthermore, as was shown in \rresolv,
perturbative $(0,2)$ vacua can remain finite with sufficiently mild
singularities in the bundle as described by reflexive or  torsion
free sheaves. In section 4 we will present a brief survey of those
sheaf theoretic methods  needed for the computation of the massless spectrum
of such  models. It will turn out, that the usual cohomology classes like 
for instance $H^1(M,V^*)$ do not determine the number of chiral 
multiplets any longer, 
instead one has to compute  what is called the global extension 
Ext$^1(M;\sV,{\cal O})$. 
Using those techniques from the theory of coherent sheaves, in section 5
one dual pair of reflexive sheaves over certain base manifolds 
will be discussed  in some detail.  
Unfortunately, the explicit calculations turn out to be much more involved
than in the bundle case, for the kernels and images of various maps
involved in the spectral sequences have to be known in detail.
For this reason, exact results for the
number of  bundle moduli are not yet available, but at least  
the general method of how to calculate them will be described.

\newsec{Review of $(0,2)$ target space duality}

It was  known for a while that at large radius completely different 
$(0,2)$ compactifications of the heterotic string can share the same
Landau-Ginzburg locus \rduala. Roughly speaking, this is possible because 
in the context of $(0,2)$ linear sigma models the non-geometric 
phases in the K\"ahler moduli space contain less information than 
the geometric phases. For instance, in a non-geometric phase
the chiral fields $P_l$ in the linear sigma model superpotential
\eqn\superpoten{S=\int d^2 z d\theta \left[ \Gamma^j W_j(X_i) + 
               P_l \Lambda^a F_a^l(X_i)  \right] }
carry a non-vanishing vacuum expectation values, implying 
the geometric complex and bundle deformations to appear on equal footing.
Consequently, an exchange of them can leave  the
superpotential invariant, whereas at large radius ($P_l=0$)
the model has drastically
changed. In \rtdual\ it was argued that the first guess, the Landau-Ginzburg
locus being a multicritical point, is not convincing and that instead
the entire moduli spaces of a dual pair seem to be isomorphic. 
 
More specifically, the quintic threefold $\IP_{4}[5]$ and a dual candidate
with base and vector bundle being the resolution of 
\eqn\quind{        V(1,1,1,2;5)\to \IP_{1,1,1,1,1,3}[4,4] }
was studied in detail. The latter model contains a phase for small radii, 
where it is described by the same Landau-Ginzburg model as the quintic.
However, for this model the space time superpotential is flat, so that
every modulus of the quintic should correspond to a modulus of the
dual $(0,2)$ model. A further confirmation of this picture was found by
calculating the dimensions of the total geometric 
moduli spaces for both models. 
Indeed in turned out that the sums of complex, K\"ahler and bundle moduli
agree, even though every individual sector is not constant.

A perturbative isomorphy of moduli spaces is what is usually called a
target space duality. Well known examples of such dualities include the
discrete $R\to 1/R$ symmetries of toroidal compactifications and mirror
symmetry in the context of Calabi-Yau compactifications. The latter one
is supposed to be correct for a much larger set of threefolds than those
exhibiting a Landau-Ginzburg phase or being  described by a
hypersurface in a weighted projective space, respectively.  
Thus, it is natural to ask whether $(0,2)$ target space duality can
be extended to models not admitting a Landau-Ginzburg phase, as well.
For deformations of $(2,2)$ models this leads to complete intersection
Calabi-Yau's (CICYs) or, more generally, to  complete intersections of
hypersurfaces in toric varieties.  

\newsec{Target space duality for CICY}

In this section we discuss several, we hope  sufficiently generic, 
examples of 
$(0,2)$ dual pairs, which do not have a Landau-Ginzburg description. 
Some familiarity with linear sigma models \rew, toric geometry \rvitja\
and homological algebra \refs{\rgrif,\rdistgr} is  assumed in the course 
of this paper. 

\subsec{A dual to $\IP_5[3,3]$}

Using the methods from \refs{\rtdual ,\rend}, we investigate the model given 
by the 
complete intersection of two hypersurfaces of weight three in  
projective space $\IP_5$. The topological data for this model
are already known \rend
\eqn\topdata{ \IP_{5}[3,3], \quad h_{21}=73,\  h_{11}=1, \ 
                          h^1(\rm{M,End}(T_M))=140, }
so that there are a total number of {\bf 214} geometric moduli. 
The linear sigma model contains two different phases. For $r>0$ there
is a Calabi-Yau phase and for $r<0$ one gets a CY/LG hybrid phase.
The hybrid phase can be described as  a fiber bundle  over $\IP_1$, where
the fiber over $(p_1,p_2)\in \IP_1$ is itself a Landau-Ginzburg model 
with superpotential
\eqn\superpot{  W=p_1 W_1(x_i) + p_2 W_2(x_i) .}
The $W_{1,2}$ denote  homogeneous polynomials  of weight three in six
coordinates $x_i$ of weight one. 
For generic points in $\IP_1$ the Landau-Ginzburg model flows in 
the infrared to a $c=6$ conformal field theory with $\chi=24$.
One peculiar feature of this Landau-Ginzburg "K3"  is its rigidness 
in the sense that
it does not have any K\"ahler deformation. In general it is not known how to
compute any further data in hybrid phases, but by using a 
trick, one can gain some pieces of information.
The "K3" Landau-Ginzburg model becomes singular over
 exactly six points in the
base, so that it is equivalent to another "K3" fibration written as
$\IP_{1,1,2,2,2,2,2}[6]$. This latter model also is the rigid
"K3"  fibered over a $\IP_1$ with six singular fibers. However, this 
model has a Landau-Ginzburg  phase and the calculation of the Hodge numbers, 
$h_{21}=73$ and $h_{11}=1$, indeed gives the desired result,
The total number
of moduli for $\IP_{1,1,2,2,2,2,2}[6]$ comes out as 244, but it is not clear 
that this really captures the number of moduli in the hybrid phase.

The base manifold of a dual $(0,2)$ model can be obtained by performing 
a (formal) conifold transition on $\IP_{5}[3,3]$. Going to a point in complex
structure moduli space, where the hypersurface equations are
\eqn\coni{  W_1=P_3(x_i), \quad\quad W_2=x_1 F_2(x_i) - x_2 G_2(x_i), }
and making a small resolution gives 
\eqn\small{\eqalign{   W_1=P_3(x_i), \quad\quad 
                       &W_2=x_1 y_1 + G_2(x_i) y_2 \cr
                       &W_3=x_2 y_1 + F_2(x_i) y_2 .\cr}}
This can be recognized as the intersection of three hypersurfaces in 
the toric variety given by the $C^*$ actions
\meno
\cl{\vbox{
\hbox{\vbox{\offinterlineskip
\def\tablespace{height2pt&\omit&&\omit&&\omit&&\omit&&\omit&&\omit&&\omit&&
                          \omit&&\omit&&\omit&&\omit&\cr}
\def\tablerule{\tablespace\noalign{\hrule}\tablespace}

\hrule\halign{&\vrule#&\strut\hskip0.2cm\hfil#\hfill\hskip0.2cm\cr
\tablespace
& $x_1$ && $x_2$ && $x_3$ && $x_4$ && $x_5$ && $x_6$ && $x_7$ && $x_8$ &&
$\Gamma_1$ && $\Gamma_2$ && $\Gamma_3$ &\cr
\tablerule
& $1$ && $0$  && $0$ && $0$ && $0$ && $0$ && $1$ && $0$  && $0$ &&
 $-1$ && $-1$ &\cr
& $1$ && $1$  && $1$ && $1$ && $1$ && $1$ &&  $0$ && $1$  &&  $-3$ &&
$-2$ && $-2$ &\cr
\tablespace}\hrule}}}}
\cl{
\hbox{{\bf Table 3.1:}{\it ~~Charges for the base}}}
\meno
Using toric methods\footnote{$^1$}{For some of the calculations involving 
toric varieties the maple packages {\it Schubert} and {\it Puntos} have 
been used \rschu} one obtains that the intersection ring on the CICY
is 
\eqn\int{ 3\eta_1^3 -3\eta_1^2\eta_2 + 3\eta_1\eta_2^2 + 
                9\eta_2^3,}
where $\eta_{1,2}$ denote the two independent sections of the 
base\footnote{$^2$}{The negative intersection numbers in \int\  appear
because  the charges in Table 3.1 are not the Mori vectors}.
The Euler number turns out to be $\chi=-120$. 
Using the algorithm to compute the cohomology classes of line bundles in 
the ambient space \refs{\rresolv,\rtdual} 
and tracing through the long exact sequences in 
bundle cohomology, one obtains for the detailed Hodge numbers 
$h_{21}=62$ and $h_{11}=2$.
The resolution of the vector bundle gives
\meno
\cl{\vbox{
\hbox{\vbox{\offinterlineskip
\def\tablespace{height2pt&\omit&&\omit&&\omit&&\omit&&\omit&&\omit&&\omit&&
                          \omit&\cr}
\def\tablerule{\tablespace\noalign{\hrule}\tablespace}

\hrule\halign{&\vrule#&\strut\hskip0.2cm\hfil#\hfill\hskip0.2cm\cr
\tablespace
& $\lambda_1$ && $\lambda_2$ && $\lambda_3$ && $\lambda_4$ && $\lambda_5$ && 
  $\lambda_1$ &&  $p_1$ && $p_2$   &\cr
\tablerule
& $1$ && $0$  && $0$ && $0$  && $0$  && $0$ && $0$ &&  $-1$  &\cr
& $0$ && $2$  && $1$ && $1$ && $1$ && $1$  &&  $-3$ && $-3$   &\cr
\tablespace}\hrule}}}}
\cl{
\hbox{{\bf Table 3.2:}{\it ~~Charges for the bundle}}}
\meno
This bundle is defined with one fermionic gauge symmetry, so that its
rank is indeed three and the resulting gauge group $E_6$. It is
given by the cohomology of the monad
\eqn\monadb{ 0\to\ {\cal O}\vert_{M} \to {\cal O}(1,0)\oplus 
                  {\cal O}(0,2)\oplus{\cal O}(0,1)^4\vert_{M}
                  \to{\cal O}(0,3)\oplus{\cal O}(1,3)\vert_{M}\to0.}
The third Chern class is $c_3(V_M)=-144$, which indeed 
comes from $h^1(M,V_M)=73$ generations and $h^2(M,V_M)=1$ antigenerations. 
The tedious computation of the bundle deformations is straightforward 
and leads to exactly $h^1(M,{\rm End}(V_M))=150$ additional moduli.
Thus, the total number of moduli is {\bf 214}, which agrees nicely 
with the $\IP_{5}[3,3]$ results. 

Now, the question is, whether one can find a phase for this $(0,2)$ model,
which coincides with the $(0,2)$ deformation of the
hybrid phase of $\IP_{5}[3,3]$.
The model under consideration has five  different phases, two of them 
geometric the other three  non-geometric. After a tedious consideration
one finds that for $r_1>0$ and $r_2<0$ the model is described
by indeed the same hybrid phase. 
Apparently, the reason is, that in non-geometric phases the $p$ fields
in the linear sigma model carry non-zero vacuum expectation value implying
that there does not exist a distinction between complex and bundle
moduli in the superpotential. 
The natural conclusion is that 
$(0,2)$ target space duality is not a special feature of
the rather restricted set of models containing a Landau-Ginzburg phase, but
carries over to the much broader class of vector bundles  over 
complete intersections in toric varieties. 

As we have seen, starting with a deformation of a $(2,2)$ model, the base 
manifold of a $(0,2)$ dual can be obtained by performing a conifold
transition on the former threefold. However, this should be regarded as a 
recipe only for getting the base of the dual, we are not claiming  that there 
is indeed a transition. This would not make sense, since the dual pair 
is supposed to be isomorphic, anyway. It simply means, that starting with a 
$(2,2)$ model and making a conifold transition, on the resolved base space  
one can define a bundle such that the former and latter model are isomorphic.

\subsec{A dual to a hypersurface in  $\IP_3\times \IP_1$}

Another type of threefolds which do not contain a pure Landau-Ginzburg
description, are hypersurfaces in products of projective spaces.
In particular, the model 
\eqn\cim{ \matrix{ \IP_3 \cr \IP_1 \cr}\left[\matrix{4 \cr 2 \cr}
          \right] }
is studied with Hodge numbers $h_{21}=86$ and $h_{11}=2$. 
The computation of bundle deformations gives $h^1(M,{\rm End}(T_M))=188$, 
so that the total number of moduli at large radius is {\bf 276}.
The model has three phases, one Calabi-Yau and two LG/CY hybrid phases.
In one of the hybrid phases one has a geometric $\IP_1$ base with homogeneous
coordinates $y_{1,2}$. Fibered over this space is the "K3", $\IP_{3}[4]$,
in its Landau-Ginzburg phase. The superpotential looks like
\eqn\superpotb{  W=\sum_{i=1}^4 p^{(2)}_i(y_1,y_2)\  x_i^4 , }
where the $p^{(2)}_i$ denote homogeneous polynomials of degree two.
Thus, one has a "K3" fibration with generically eight singular fibers.
One encounters the same situation for the K3 fibration $\IP_{1,1,2,2,2}[8]$,
which consistently has Hodge numbers $(h_{21},h_{11})=(86,2)$, as well.
The total number of moduli for the latter model is 294, but as above it
is not clear whether this really counts  the number of moduli seen in  
the hybrid phase. The second hybrid phase is a discrete $\ZZ_2$ bundle
over $\IP_3$. On the boundary of these two hybrid phases one finds a 
gauged Landau-Ginzburg phase. 

The data of the $(0,2)$ dual model can be obtained straightforwardly.
The base is the threefold defined by the $C^*$ actions in Table 3.3

\meno
\cl{\vbox{
\hbox{\vbox{\offinterlineskip
\def\tablespace{height2pt&\omit&&\omit&&\omit&&\omit&&\omit&&\omit&&\omit&&
                          \omit&&\omit&&\omit&\cr}
\def\tablerule{\tablespace\noalign{\hrule}\tablespace}

\hrule\halign{&\vrule#&\strut\hskip0.2cm\hfil#\hfill\hskip0.2cm\cr
\tablespace
& $x_1$ && $x_2$ && $x_3$ && $x_4$ && $x_5$ && $x_6$ && $x_7$ && $x_8$ &&
$\Gamma_1$ && $\Gamma_2$ &\cr
\tablerule
& $2$ && $1$  && $1$ && $1$ && $1$ && $0$ && $0$ && $0$  && $-3$ &&
 $-3$ &\cr
& $2$ && $0$  && $0$ && $0$ && $0$ && $1$ &&  $1$ && $0$  &&  $-2$ &&
$-2$  &\cr
& $1$ && $0$  && $0$ && $0$ && $0$ && $0$ &&  $0$ && $1$  &&  $-1$ &&
$-1$  &\cr
\tablespace}\hrule}}}}
\cl{
\hbox{{\bf Table 3.3:}{\it ~~Charges for the base}}}
\meno
Computing the intersection ring for the toric variety gives
\eqn\intb{ 2\eta_1^3 + 4\eta_1^2\eta_2 + \eta_1\eta_2\eta_3 - 
        2\eta_1\eta_3^2 - 2\eta_2\eta_3^2 + 8\eta_3^3,}
which allows one to calculate the Euler number $\chi=-144$. 
The more refined cohomology calculation reveals $(h_{21},h_{11})=(75,3)$.
The bundle on the threefold is described by the data in Table 3.4
\meno
\cl{\vbox{
\hbox{\vbox{\offinterlineskip
\def\tablespace{height2pt&\omit&&\omit&&\omit&&\omit&&\omit&&\omit&&
                          \omit&\cr}
\def\tablerule{\tablespace\noalign{\hrule}\tablespace}

\hrule\halign{&\vrule#&\strut\hskip0.2cm\hfil#\hfill\hskip0.2cm\cr
\tablespace
& $\lambda_1$ && $\lambda_2$ && $\lambda_3$ && $\lambda_4$ && $\lambda_5$ && 
  $\lambda_6$ &&   $p$   &\cr
\tablerule
& $0$ && $1$  && $1$ && $0$  && $0$  && $2$ &&  $-4$  &\cr
& $0$ && $0$  && $0$ && $1$ && $1$ && $0$  &&  $-2$     &\cr
& $1$ && $0$  && $0$ && $0$ && $0$ && $0$  &&  $-1$     &\cr
\tablespace}\hrule}}}}
\cl{
\hbox{{\bf Table 3.4:}{\it ~~Charges for the bundle}}}
\meno
The bundle is defined with two fermionic gauge symmetries as the 
cohomology of the monad
\eqn\monadb{ 0\to\ {\cal O}^2\vert_{M} \to {\cal O}(0,0,1)\oplus 
                  {\cal O}(1,0,0)^2\oplus{\cal O}(0,1,0)^2
                  \oplus{\cal O}(2,0,0)\vert_{M}
                  \to{\cal O}(4,2,1)\vert_{M}\to0.}
The bundle valued cohomology computation gives $h^1(M,V_M)=86$, $h^2(M,V_M)=2$
and $h^1(M,{\rm End}(V_M))=198$, so that the number of moduli indeed adds up 
to {\bf 276}. 
This $(0,2)$ model has six different phases, three of them non-geometric
and the sector spanned by the vertices
\eqn\verta{ \left\{ (-4,-2,-1),(0,1,0),(0,0,1) \right\} }
in the secondary fan, contains a phase which is the same CY/LG hybrid 
\superpotb\ as
for the $(0,2)$ deformation of the original model \cim. It can be
checked that the second  CY/LG hybrid phase of \cim\  corresponds to
the sector
\eqn\vertb{ \left\{ (-4,-2,-1),(1,0,0),(0,0,1) \right\} }
in the K\"ahler moduli space and that on the boundary one gets a gauged
Landau-Ginzburg phase.

\subsec{A strict $(0,2)$ dual pair }

So far, solely dual pairs with one model
being the deformation of a $(2,2)$ model have been studied. In these cases one
new K\"ahler class is introduced in the base. We further pursue 
a model which has been introduced in \rdualb\ and lives on the base
ambient space $\IP_3\times \IP_2$. 
Model A is defined by the data in Table 3.5 for the base
\meno
\cl{\vbox{
\hbox{\vbox{\offinterlineskip
\def\tablespace{height2pt&\omit&&\omit&&\omit&&\omit&&\omit&&\omit&&
                          \omit&&\omit&&\omit&\cr}
\def\tablerule{\tablespace\noalign{\hrule}\tablespace}

\hrule\halign{&\vrule#&\strut\hskip0.2cm\hfil#\hfill\hskip0.2cm\cr
\tablespace
& $x_1$ && $x_2$ && $x_3$ && $x_4$ && $x_5$ && $x_6$ && $x_7$  &&
$\Gamma_1$ && $\Gamma_2$ &\cr
\tablerule
& $1$ && $1$  && $1$ && $1$ &&  $0$ && $0$ && $0$  && $-2$ &&
 $-2$ &\cr
& $0$ && $0$  && $0$ && $0$ && $1$ && $1$ &&  $1$  &&  $-2$ &&
$-1$  &\cr
\tablespace}\hrule}}}}
\cl{
\hbox{{\bf Table 3.5:}{\it ~~Charges for the base}}}
and the data in Table 3.6 for the vector bundle 
\meno
\cl{\vbox{
\hbox{\vbox{\offinterlineskip
\def\tablespace{height2pt&\omit&&\omit&&\omit&&\omit&&
                          \omit&\cr}
\def\tablerule{\tablespace\noalign{\hrule}\tablespace}

\hrule\halign{&\vrule#&\strut\hskip0.2cm\hfil#\hfill\hskip0.2cm\cr
\tablespace
& $\lambda_1$ && $\lambda_2$ && $\lambda_3$ && $\lambda_4$ &&   $p$   &\cr
\tablerule
& $0$ && $0$  && $1$ && $2$  && $-3$  &\cr
& $1$ && $1$  && $0$ && $0$ &&  $-2$     &\cr
\tablespace}\hrule}}}}
\cl{
\hbox{{\bf Table 3.6:}{\it ~~Charges for the bundle}}}
\meno
This model is consistent without any fermionic gauge symmetry, so that
the rank three bundle is simply given by the exact sequence
\eqn\exseq{ 0\to V_M\to {\cal O}(0,1)^2\oplus 
                  {\cal O}(1,0)\oplus{\cal O}(2,0)\vert_{M}
                  \to {\cal O}(3,2)\vert_{M}\to 0}
shortening the bundle cohomology calculations considerably.
One gets $h^1(M,V_M)=84$ generations in the {\bf 27} representation of $E_6$
and no antigeneration. Adding up the Hodge numbers of the base,
$h_{21}=62$ and $h_{11}=2$, and the bundle deformations, 
$h^1(M,{\rm End}(V_M))=184$ one obtains the dimension of the total moduli 
space {\bf 248}. 

After exchanging $\{ W_1, W_2\} \leftrightarrow \{F_1,F_4\}$ the data
for the base of the dual model B are
\meno
\cl{\vbox{
\hbox{\vbox{\offinterlineskip
\def\tablespace{height2pt&\omit&&\omit&&\omit&&\omit&&\omit&&\omit&&
                          \omit&&\omit&&\omit&\cr}
\def\tablerule{\tablespace\noalign{\hrule}\tablespace}

\hrule\halign{&\vrule#&\strut\hskip0.2cm\hfil#\hfill\hskip0.2cm\cr
\tablespace
& $x_1$ && $x_2$ && $x_3$ && $x_4$ && $x_5$ && $x_6$ && $x_7$  &&
$\Gamma_1$ && $\Gamma_2$ &\cr
\tablerule
& $1$ && $1$  && $1$ && $1$ &&  $0$ && $0$ && $0$  && $-3$ &&
 $-1$ &\cr
& $0$ && $0$  && $0$ && $0$ && $1$ && $1$ &&  $1$  &&  $-1$ &&
$-2$  &\cr
\tablespace}\hrule}}}}
\cl{
\hbox{{\bf Table 3.7:}{\it ~~Charges for the base}}}
The vector bundle is defined by the charges in Table 3.8
\meno
\cl{\vbox{
\hbox{\vbox{\offinterlineskip
\def\tablespace{height2pt&\omit&&\omit&&\omit&&\omit&&
                          \omit&\cr}
\def\tablerule{\tablespace\noalign{\hrule}\tablespace}

\hrule\halign{&\vrule#&\strut\hskip0.2cm\hfil#\hfill\hskip0.2cm\cr
\tablespace
& $\lambda_1$ && $\lambda_2$ && $\lambda_3$ && $\lambda_4$ &&   $p$   &\cr
\tablerule
& $0$ && $1$  && $1$ && $1$  && $-3$  &\cr
& $1$ && $0$  && $0$ && $1$ &&  $-2$     &\cr
\tablespace}\hrule}}}}
\cl{
\hbox{{\bf Table 3.8:}{\it ~~Charges for the bundle}}}
\meno
The charged matter spectrum turns out to be same as for model A, 
$h^1(M,V_M)=84$ and $h^2(M,V_M)=0$. The neutral matter receives contributions
from $h_{21}=59$ complex, $h_{11}=2$ K\"ahler and $h^1(M,{\rm End}(V_M))=
187$ bundle deformations. As expected, the total number of moduli, {\bf 248}, 
is identical to the result for model A. 
One can further show, that model A and model B are identical in their
two CY/LG hybrid phases. 

We hope, that the above examples have convinced the reader that
$(0,2)$ target space duality is a general pattern in the class of 
$(0,2)$ models
and not a rare exception for those models allowing a Landau-Ginzburg
description. It should be clear, that something deep in mathematics 
is going on here. There  should exist a duality map
acting on vector bundles  over toric varieties so that the sum
$h_{21}+h_{11}+h^1(M,{\rm End(V)})$ is invariant. This is very similar
to mirror symmetry for $(2,2)$ models, where $h_{21}+h_{11}$ is 
constant. The right way to address these question in mathematical terms
is probably to find a combinatoric description of coherent sheaves
over toric varieties which also includes the bundle deformations.

\newsec{Reflexive sheaves}

In \rresolv\ it was nicely shown in the framework of linear sigma
models that $(0,2)$ models can live with some mild singularities 
in the bundle. In particular, the defining maps $F_a$ in a sequence 
\eqn\exseqc{ 0\to V_M\to E_M 
            \buildrel F_a \over \to {\cal O}(D)\vert_{M}\to 0}
are allowed to vanish on a codimension three locus $S$  in the
threefold $M$. In \exseqc\ $E_M$ denotes a vector bundle and ${\cal O}(D)$ 
the line bundle associated to the divisor $D\subset M$. 
It was shown that the parameter space of the linear sigma model remains
compact by gluing in some proturberances at the singularities.
As a consequence, in the large radius limit the sequence  \exseqc\
is no longer exact, but can be extended to an exact sequence by including
the cokernel of the map $F_a$
\eqn\exseqd{ 0\to V_M\to E_M 
            \buildrel F_a \over \to {\cal O}(D)\vert_{M}\to 
            {\cal O}(D)\vert_{S} \to 0.}
Thus, the coherent sheaf $V_M$ fails to be locally free
over a sublocus of codimension three in the threefold $M$.  
For the cases studied in this paper, the sublocus $S$ is describable as 
the complete intersection of three hypersurfaces $S=\{ f_1=f_2=f_3=0\}$.
The ideal generated by these $f_i$s is denoted as $I$. 
Then ${\cal O}(D)\vert_{S}$ fits into the exact sequence
\eqn\ideal{ 0\to I\otimes {\cal O}(D)\vert_{M} \to {\cal O}(D)\vert_{M}\to
            {\cal O}(D)\vert_{S}\to 0. }
As explained in the following subsection, for sheaves of  type \exseqd\ 
reflexivity still holds in the sense $V^{**}_M=V_M$.
A sheaf like ${\cal O}(D)\vert_{S}$ supported only at a finite number 
of points is called a skyscraper sheaf.
Many of the salient features of vector bundles do not generalize trivially
to coherent sheaves but there exists a nice mathematical theory 
describing   the peculiarities occurring  for this latter  structure.
For mathematical details of the following brief digression on 
coherent sheaves the reader is referred to the existing literature \rgrif.
A good introduction to sheaf theory for physicist has been presented 
in \rdistgr. 
We do not repeat everything mentioned in \rdistgr, but continue their
introduction and focus  on those technical tools needed for the practical 
purpose of calculating the massless spectrum of $(0,2)$ string models.  
A less extensive survey on coherent sheaves has also been provided
recently in \rsharpe. 

\subsec{Coherent sheaves}

A sheaf $\sf$ of ${\cal O}$ modules over a complex manifold $M$ is said to be
{\it coherent}, if locally it has a presentation
\eqn\pres{  {\cal O}^{(p)}\to {\cal O}^{(q)}\to \sf\to 0 .}
The crucial property for the following is that a coherent  sheaf ${\cal F}$ 
always allows what  is called a {\it global syzygy} or a {\it locally free
resolution}, 
\eqn\reso{ E.(\sf):  0\buildrel \delta \over \to E_n \buildrel \delta \over
       \to \ldots \buildrel \delta \over \to E_1 \buildrel \delta \over
       \to \sf\to 0,}
where all sheaves $E_j$ are locally free (vector bundles) and the sequence
is exact.
For vector bundles one is used to the fact, that tensoring with another vector
bundle or taking
the dual of a short exact sequence yields another short exact sequence.
This is not any longer true for coherent sheaves. The way in which it fails
is measured by the sheaves {\it (local) extension}, 
$\underline{\it Ext}^n(\sf,\sg)$, and
{\it (local) torsion}, $\underline{\it Tor}_n(\sf,\sg)$,
defined as the cohomology and homology
\eqn\exttor{\eqalign{ \underline{\it Ext}^n(\sf,\sg)=H^n_{\delta}\left(
                       Hom(E.(\sf),\sg) \right) \cr
                      \underline{\it Tor}_n(\sf,\sg)=H_n^{\delta}\left(
                      E.(\sf)\otimes \sg \right), \cr}}
respectively. From \reso\ it is obvious that 
$\underline{\it Ext}^0(\sf,\sg)=Hom(\sf,\sg)$ 
and $\underline{\it Tor}_0(\sf,\sg)=\sf\otimes \sg$. 
Moreover, as one is used to from ordinary cohomology, short  exact sequences 
of sheaves
\eqn\shortseq{ 0 \to \sf \to \sg \to \sh \to 0}
imply long exact sequences of  $\underline{\it Ext}^n$ and 
$\underline{\it Tor}_n$. For instance, the long exact sequence for
local extension looks like
\eqn\longes{ 0 \to Hom(\sf,\sm) \to  Hom(\sg,\sm) \to  Hom(\sh,\sm) \to 
               \underline{\it Ext}^1(\sf,\sm) \to 
               \underline{\it Ext}^1(\sg,\sm) \to \ldots }
showing  in which sense  (local)  $\underline{\it Ext}$ measures the extend
to which $Hom(\, .\, ,\sm)$ fails to be exact. 
Since for a locally free sheaf one can choose $E_0=\sf$ with
all other bundles $E_i$s vanishing, it is clear that 
the higher extensions $\underline{\it Ext}^n(\sf,\sg)$ vanish for
$n>0$, so that one recovers the familiar features of  vector bundles. 

As an application of working with local extensions it is shown that sheaves
defined by an exact sequence like \exseqd\ are indeed reflexive.
Joining the sequences \exseqd\ and \ideal\ into the diagram
\vfill\eject
\phantom{ff}
\vskip -3cm
\eqn\diagrama{ 0\to V_M\to E_M \to
              \vcenter{\vskip 2.2cm \hbox{0}
                     \hbox{$\downarrow$}
                     \hbox{$I\otimes {\cal O}(D)\vert_{M}\to 0$}
                     \hbox{$\downarrow$}
                     \hbox{${\cal O}(D)\vert_{M}$}
                     \hbox{$\downarrow$}
                     \hbox{${\cal O}(D)\vert_{S}$}
                     \hbox{$\downarrow$}
                     \hbox{0} }
                      }
and applying the $\underline{\it Ext}^q(\, .\, ,{\cal O})$ functor leads to 
the diagram
\vskip 1.5cm
\eqn\diagramb{ 0\to 
              \vcenter{\vskip -1.25cm 
                \hbox{$\underline{\it Ext}^2({\cal O}(D)\vert_{S},{\cal O})$}
                \hbox{$\uparrow$}
                \hbox{$\underline{\it Ext}^1(I\otimes {\cal O}(D)\vert_{M}
                                               ,{\cal O})$}
                \hbox{$\uparrow$}
                \hbox{$\underline{\it Ext}^1({\cal O}(D)\vert_{M}
                                               ,{\cal O})$}
                \hbox{$\uparrow$}
                \hbox{$\underline{\it Ext}^1({\cal O}(D)\vert_{S},{\cal O})$}
                \hbox{$\uparrow$}
                \hbox{${\it Hom}(I\otimes {\cal O}(D)\vert_{M}
                                               ,{\cal O}) \to E^*_M \to 
              V^*_M \to\underline{\it Ext}^1(I\otimes {\cal O}(D)\vert_{M}
                                               ,{\cal O})\to 0$}
                \hbox{$\uparrow$}
                \hbox{${\it Hom}({\cal O}(D)\vert_{M}
                                               ,{\cal O})$}
                \hbox{$\uparrow$}
                \hbox{${\it Hom}({\cal O}(D)\vert_{S},{\cal O})$}
                \hbox{$\uparrow$}
                     \hbox{0} } }
Using the Koszul resolution, it is proven in \rgrif\  
that if $S$ has codimension $n$, one gets for the local extensions
\eqn\extq{ \underline{\it Ext}^q({\cal O}_S,{\cal O})=0\quad\quad
{\rm for} \  q\in\{0,\ldots,n-1\} .}
Using this for our codimension three locus and that for locally free 
sheaves $\underline{\it Ext}^q$ vanishes
for $q>0$ the diagram \diagramb\ collapses to the short exact 
sequence\footnote{$^1$}{Note, that a map between coherent sheaves can be 
injective without being injective on each fiber}
\eqn\diagramc{ 0\to {\cal O}^*(D)\vert_{M} 
               \to E^*_M \to V^*_M \to 0.} 
Dualizing \diagramc\ and
using reflexivity of locally free sheaves leads 
us back to the sequence 
\eqn\diagramc{ 0\to V^{**}_M \to E_M \to {\cal O}(D)\vert_{M}  \to
               {\cal O}(D)\vert_{S} \to 0,}
implying the desired result $V^{**}_M=V_M$.

One essential property of vector bundles over compact, complex
manifolds is Serre duality
\eqn\serre{  H^p(M,V)^*\cong H^{n-p}(M,V^*\otimes {\cal K}_M)}
with ${\cal K}_M$ denoting the canonical bundle of $M$.
The bundle valued cohomologies are understood as global \v{C}ech cohomologies.
In order to formulate the sheaf theoretic generalization one introduces
so called global Ext. 
It is defined as the hypercohomology of the complex of sheaves 
$Hom(E.(\sf),\sg)$ over $M$:
\eqn\globalext{ {\rm Ext}(M;\sf,\sg)=\HH^*(M;Hom\left(E.(\sf),\sg)\right) }
In general,
the hypercohomolgy of a complex of sheaves ${\cal K}=({\cal K}^*,d)$ over  a
manifold $M$ is defined as the cohomology of the associated 
single complex $(C^*(\underline{U},{\cal K}),D=\delta+d)$ of
the double complex 
\eqn\doublec{\left\{ C^p(\underline{U},{\cal K}^q);
           \delta,d\right\}. }
Here  $C^p(\underline{U},{\cal K}^q)$ denotes the \v{C}ech cochains of degree
$p$ with values in the sheaf ${\cal K}^q$ and $\delta$ is the \v{C}ech 
coboundary operator. 
One of the two abuting spectral sequences of the double complex
\doublec\ leads after two steps to 
\eqn\twosteps{\eqalign{  &E_2^{p,q}=H^p(M,\underline{\it Ext}^q(\sf,\sg)) \cr
                &E_\infty^{p,q}\Rightarrow {\rm Ext}^{p+q}(M;\sf,\sg ). }}
Now, it has become evident  that for $\sf$ a locally free sheaf, 
one obtains that global Ext reduces to the ordinary  cohomology  groups
\eqn\lfs{ {\rm Ext}^q(M,\sf,\sg)\cong H^q(M,\sf^*\otimes \sg) .}
In particular for the structure sheaf ${\cal O}$ \lfs\ means
${\rm Ext}^q(M,{\cal O},\sg)\cong H^q(M, \sg)$. 

What is needed in the following section is a tool of how to calculate 
global Ext$(M;\sf,\sg)$ for a sheaf $\sf$ fitting into an exact sequence 
of sheaves 
\eqn\essheaf{ 0\to \sh_n \buildrel \eta\over \to \sh_{n-1}\buildrel \eta\over
       \to\ldots\buildrel \eta\over \to\sh_0\to \sf\to 0}
where for each individual $\sh_i$ global Ext is known. Note, that it is  not
required that  the $\sh_i$s are locally free.  
As usual in homological algebra one considers
the double complex $\left\{ C^p(\underline{U},Hom(\sh_q,\sg));
           D,\eta\right\}$
and running the two abutting spectral sequences allows one to
express global Ext$(M;\sf,\sg)$ in terms of Ext$(M;\sh_q,\sg)$. 
One explicit example will be discussed in the following section.

Finally, Serre duality for a general coherent sheaf becomes
\eqn\sersheaf{  H^p(M,\sf)^*\cong {\rm Ext}^{n-p}(M;\sf,{\cal K}_M ).}
This closes the compact digression on coherent sheaves, in which 
the necessary technical tools for dealing with 
reflexive sheaves of the kind \exseqd\  have been provided. 

\subsec{Example of a reflexive sheaf}

In this subsection the methods introduced in the
last subsection will be applied to the determination of the charged 
massless spectrum of a specific $(0,2)$ model. 
Before that  the relation between  the
different massless modes of the string theory and the various
cohomology groups of the sheaf over the Calabi-Yau manifold $M$ has
to be clarified. 
For a vector bundle $V$ the number of generations and antigenerations
were determined by $H^1(M,V)$ and $H^1(M,V^*)$. Furthermore, some
of the uncharged chiral fields corresponded to the traceless part
of $H^1(M,Hom(V,V))$. In view of \lfs\ and Serre duality \sersheaf\
the generalization to a coherent sheaf $\sV$  appears to be that
${\rm Ext}^1(M;{\cal O},\sV)$ and
${\rm Ext}^1(M;\sV,{\cal O})$ count the generations and antigenerations,
respectively. Then it seems natural, that the part of the number of uncharged
singlets is related to  ${\rm Ext}^1(M;\sV,\sV)$.

Consider the following two singular configurations
\eqn\tdualm{\eqalign{ &A:\  V(1,1,2,4;8)\to \IP_{1,1,1,3,3,3}[6,6] \cr
                      &B:\  V(1,1,2,2,2;8)\to \IP_{1,1,1,3,3,3}[8,4] ,\cr }}
the resolution of which is supposed to lead to a dual pair of $(0,2)$ 
models. Both models share the same Landau Ginzburg locus, for which 
one obtains $N_{27}=96$ generations and $N_{\overline{27}}=4$ antigenerations.
The resolution of the base manifold of model A gives
\meno
\cl{\vbox{
\hbox{\vbox{\offinterlineskip
\def\tablespace{height2pt&\omit&&\omit&&\omit&&\omit&&\omit&&\omit&&
                          \omit&&\omit&&\omit&\cr}
\def\tablerule{\tablespace\noalign{\hrule}\tablespace}

\hrule\halign{&\vrule#&\strut\hskip0.2cm\hfil#\hfill\hskip0.2cm\cr
\tablespace
& $x_1$ && $x_2$ && $x_3$ && $x_4$ && $x_5$ && $x_6$ && $x_7$  &&
$\Gamma_1$ && $\Gamma_2$ &\cr
\tablerule
& $1$ && $1$  && $1$ && $0$ &&  $0$ && $1$ && $0$   && $-2$ &&
 $-2$ &\cr
& $3$ && $3$  && $3$ && $1$ &&  $1$ && $0$ && $1$   && $-6$ &&
 $-6$ &\cr
\tablespace}\hrule}}}}
\cl{
\hbox{{\bf Table 4.1:}{\it ~~Charges for the base}}}
with Euler number $\chi=-144$ resulting from $h_{21}=77$  complex deformations
and $h_{11}=5$  K\"ahler deformations. A possible resolution of the bundle 
is given by the data in Table 4.2
\meno
\cl{\vbox{
\hbox{\vbox{\offinterlineskip
\def\tablespace{height2pt&\omit&&\omit&&\omit&&\omit&&
                          \omit&\cr}
\def\tablerule{\tablespace\noalign{\hrule}\tablespace}

\hrule\halign{&\vrule#&\strut\hskip0.2cm\hfil#\hfill\hskip0.2cm\cr
\tablespace
& $\lambda_1$ && $\lambda_2$ && $\lambda_3$ && $\lambda_4$ &&
  $p$   &\cr
\tablerule
& $0$ && $1$  && $0$ && $2$  && $-3$  &\cr
& $1$ && $1$  && $2$ && $4$  &&  $-8$     &\cr
\tablespace}\hrule}}}}
\cl{
\hbox{{\bf Table 4.2:}{\it ~~Charges for the sheaf}}}
\meno
without any  fermionic gauge symmetry. 
Naively, the third Chern class of the bundle comes out as $c_3(V_M)=-192$
which is not what one expects from the Landau-Ginzburg computation. 
The reason for this mismatch is that $V$ actually is not
a bundle but a reflexive sheaf. This can be seen by looking
at the functions $F_a$ in more detail. Since they have to be of the form
\eqn\funcfa{\eqalign{ &F_1=x_6\, p_{(2,7)}, \quad F_2=p_{(2,7)} \cr
              &F_3=x_6\, p_{(2,6)}, \quad F_4=p_{(1,4)},\cr}}
they simultaneously  vanish on the complete intersection of the three divisors
$S=\{ x_6=p_{(2,7)}=p_{(1,4)}=0 \}\subset M$. The set $S$ consists of exactly 
four points.
More formally this can be seen by using the Koszul sequence for the 
structure sheaf on $S$
\eqn\koszul{\eqalign{&0\to {\cal O}(-4,-11)\vert_M \to {\cal O}(-3,-11)\oplus
            {\cal O}(-3,-7)\oplus{\cal O}(-2,-4)\vert_M  \to \cr
           &\to{\cal O}(-2,-7)\oplus{\cal O}(-1,0)\oplus
        {\cal O}(-1,-4)\vert_M \to {\cal O}\vert_M \to {\cal O}_S \to 0 \cr}}
to compute $h^0(M,{\cal O}_S)=4$. 

It is worth mentioning, that in the Calabi-Yau phase of the linear 
sigma $\{ r_1>0,r_2>3r_1\}$ 
vanishing of the D-terms 
\eqn\dterms{\eqalign{  &\vert x_1\vert^2 + \vert x_2\vert^2+
                          \vert x_3\vert^2+\vert x_6\vert^2 -3
                    \vert p\vert^2 =r_1 \cr
                     &\vert x_4\vert^2 + \vert x_5\vert^2+
                          \vert x_7\vert^2 -3\vert x_6\vert^2 +
                    \vert p\vert^2 =r_2-3r_1 \cr}}
still forces the parameter space to be compact. Over the singular set
$S$ the field $p$ is no longer set to zero, but partly parameterizes
four new $\IP_1$ proturberances glued in automatically to resolve the 
singularity. 
Thus, as already observed in \rresolv\ such mild singularities in the 
vector bundle do not lead to singularities in the conformal field theory and 
string theory is  able to resolve them even on the perturbative level. 

The reflexive sheaf ${\cal V}_M$ is defined via the exact sequence
\eqn\exseq{0\to {\cal V}_M \to {\cal O}(0,1)\oplus 
                  {\cal O}(1,1)\oplus{\cal O}(0,2)\oplus{\cal O}(2,4)\vert_{M}
              \to{\cal O}(3,8)\vert_{M}\to {\cal O}(3,8)\vert_{S}\to 0.}
In order to determine the number of generations
Ext$^1(M;{\cal O},\sV_M)$ and antigenerations Ext$^1(M;\sV_M,{\cal O})$ one
has to run  the spectral sequence. 
After determining all the intermediate cohomology
classes, one finally arrives at the spectral sequence  
\vskip 0.1in
\meno
\cl{\vbox{
\hbox{\vbox{\offinterlineskip
\def\tablespace{height2pt&\omit&&\omit&&\omit&&\omit&&
                          \omit&\cr}
\def\tablerule{\tablespace\noalign{\hrule}\tablespace}

\hrule\halign{&\vrule#&\strut\hskip0.2cm\hfil#\hfill\hskip0.2cm\cr
\tablespace
& $$ && ${\cal V}_{M}$ && $\bigoplus {\cal O}(m,n)\vert_{M}$ && 
   ${\cal O}(3,8)\vert_{M}$ && ${\cal O}(3,8)\vert_{S}$ &\cr
\tablerule
& Ext$^0(M;{\cal O},.)$ && $0$  && $36$ && $132$ &\omit\hskip -4pt $\buildrel 
  \alpha\over \to$ & $4$    &\cr
& Ext$^1(M;{\cal O},.)$ && $96$  && $0$ && $0$ && $0$  &\cr
& Ext$^2(M;{\cal O},.)$ && $4$  && $0$ && $0$  && $0$    &\cr
& Ext$^3(M;{\cal O},.)$ && $0$  && $0$ && $0$ && $0$  &\cr
\tablespace}\hrule}}}}
\cl{
\hbox{{\bf Table 4.3:}{\it ~~Spectral sequence for determining 
                             Ext$^1(M;{\cal O},\sV_M)$ }}}
Here, it has been used  that the image of the map $\alpha$ has dimension zero,
a fact which can be seen simply by observing that every section
of ${\cal O}(3,8)\vert_{M}$ must contain at least one $x_6$ and thus
vanishes when restricted to $S$. As the first column in Table 4.3 shows,
the number of charged chiral multiplets agrees with the Landau-Ginzburg
result.

As already mentioned, generally for reflexive sheaves one is forced to study 
some of the maps involved in the sequences in detail. This makes live much
harder than in the bundle case, where it often suffices to know merely the
dimensions of the cohomology groups without the action of the various
maps. 
Even though the number of antigenerations Ext$^1(M;\sV,{\cal O})$ is 
determined
by Serre duality, let us verify it explicitly by using the formalism
of coherent sheaves. By using Ext$^q(M;{\cal L},{\cal O})=H^q(M,{\cal L}^*)$ 
for the line bundles involved and  Ext$^q(M;{\cal O}(3,8)\vert_{S},{\cal O})=
{\rm Ext}^{3-q}(M;{\cal O},{\cal O}(3,8)\vert_{S})$ for the skycraper sheaf,
one obtains the following spectral sequence
\vskip 0.1in
\meno
\cl{\vbox{
\hbox{\vbox{\offinterlineskip
\def\tablespace{height2pt&\omit&&\omit&&\omit&&\omit&&
                          \omit&\cr}
\def\tablerule{\tablespace\noalign{\hrule}\tablespace}

\hrule\halign{&\vrule#&\strut\hskip0.2cm\hfil#\hfill\hskip0.2cm\cr
\tablespace
& $$ && ${\cal O}(3,8)\vert_{S}$ && ${\cal O}(3,8)\vert_{M}$ &&
  $\bigoplus {\cal O}(m,n)\vert_{M}$ && ${\cal V}_{M}$ &\cr
\tablerule
& Ext$^0(M;.,{\cal O})$ && $0$  && $0$ && $0$  && $0$    &\cr
& Ext$^1(M;.,{\cal O})$ && $0$  && $0$ && $0$ && $4$  &\cr
& Ext$^2(M;.,{\cal O})$ && $0$  && $0$ && $0$  && $96$    &\cr
& Ext$^3(M;.,{\cal O})$ && $4$  &\omit\hskip -4pt $\buildrel \alpha^t\over \to$
                                 & $132$ && $36$ && $0$  &\cr
\tablespace}\hrule}}}}
\cl{
\hbox{{\bf Table 4.4:}{\it ~~Spectral sequence for determining 
                             Ext$^1(M;\sV_M,{\cal O})$ }}}

Note again, that this is definitely different from $H^q(M,\sV^*_M)$ with 
$(h^0,\ldots,h^3)=(0,0,96,0)$. 
Due to the necessity of determining the kernels and images of various
maps, the computation of Ext$^q(M;\sV_M,\sV_M)$ is fairly involved, even though 
in principal, the algorithm described in \rtdual\ for vector bundles carries
over as long as one carefully works with extensions instead of simply
with bundle valued cohomologies. 

The dual model B can be treated completely analogously. 
The resolution of the base manifold gives
\meno
\cl{\vbox{
\hbox{\vbox{\offinterlineskip
\def\tablespace{height2pt&\omit&&\omit&&\omit&&\omit&&\omit&&\omit&&
                          \omit&&\omit&&\omit&\cr}
\def\tablerule{\tablespace\noalign{\hrule}\tablespace}

\hrule\halign{&\vrule#&\strut\hskip0.2cm\hfil#\hfill\hskip0.2cm\cr
\tablespace
& $x_1$ && $x_2$ && $x_3$ && $x_4$ && $x_5$ && $x_6$ && $x_7$  &&
$\Gamma_1$ && $\Gamma_2$ &\cr
\tablerule
& $1$ && $1$  && $1$ && $0$ &&  $0$ && $1$ && $0$   && $-3$ &&
 $-1$ &\cr
& $3$ && $3$  && $3$ && $1$ &&  $1$ && $0$ && $1$   && $-8$ &&
 $-4$ &\cr
\tablespace}\hrule}}}}
\cl{
\hbox{{\bf Table 4.5:}{\it ~~Charges for the base}}}
with Euler number $\chi=-208$ coming from $h_{21}=109$  complex deformations
and $h_{11}=5$  K\"ahler deformations. A possible resolution of the bundle 
is given by the data in Table 4.6
\meno
\cl{\vbox{
\hbox{\vbox{\offinterlineskip
\def\tablespace{height2pt&\omit&&\omit&&\omit&&\omit&&\omit&&
                          \omit&\cr}
\def\tablerule{\tablespace\noalign{\hrule}\tablespace}

\hrule\halign{&\vrule#&\strut\hskip0.2cm\hfil#\hfill\hskip0.2cm\cr
\tablespace
& $\lambda_1$ && $\lambda_2$ && $\lambda_3$ && $\lambda_4$ &&  $\lambda_5$ && 
  $p$   &\cr
\tablerule
& $0$ && $1$  && $0$ && $1$ && $1$  && $-3$  &\cr
& $1$ && $1$  && $2$ && $2$ && $2$   &&  $-8$     &\cr
\tablespace}\hrule}}}}
\cl{
\hbox{{\bf Table 4.6:}{\it ~~Charges for the sheaf}}}
\meno
with one fermionic gauge symmetry. 
Again the bundle is singular over the set $S$ consisting of four points 
leading again  to
a reflexive sheaf. The generations and antigenerations turn out to be
identical to model A, providing some evidence that model A and B are
in fact dual to each other.

\newsec{Conclusion and Outlook}

In this paper, by studying some generic examples
it has been argued
that the existence of a Landau-Ginzburg phase is not
essentiell for $(0,2)$ target space duality. 
Instead it suffices that a potentially dual pair shares some
non-geometric locus in the extended K\"ahler moduli space. 
For  a couple of such dual pairs it has been shown that
even in the geometric large radius phase the  dimensions
of the total moduli spaces agree. 
Furthermore, a brief survey of the adequate mathematical formalism
for dealing with more general coherent sheaves was presented. 
Using these methods to compute part of the massless spectrum
some evidence was provided, that relaxing vector bundles 
to reflexive sheaves does not change the duality picture at all. 
The painstaking task of computing the exact number of sheaf 
deformations has to await a more ambitious attempt. 

It would also be interesting to investigate what the F-theory dual
picture is for this heterotic duality. Apparently, the elliptic
fibers of the heterotic threefolds map under the duality transformation
as
\eqn\fiber{ \IP_{1,2,3}[6] \to \IP_{1,1,2}[4] \to \IP_{1,1,1}[3] 
          \to \IP_{1,1,1,1}[2,2] . }
Momentarily, it is not even known what the F-dual fourfold for 
the last three elliptic fibers in \fiber\ are.

\bigbreak\bigskip\bigskip\centerline{{\bf Acknowledgments}}\nobreak

I would like to thank Peter Mayr and Andreas Wi\ss kirchen for discussion.
This  work is supported by NSF grant PHY--9513835.

\vskip 3cm 

\listrefs
\bye